\documentclass[preprint,showpacs,showkeys,aps,prc]{revtex4-1}
\usepackage{graphicx}% Include figure files
\usepackage[caption = false]{subfig}
\usepackage{dcolumn}% Align table columns on decimal point
\usepackage{bm}%

\begin{document}

\title{Deformation of a magnetized neutron star}

\date{\today}

\author{Ritam Mallick} 
\email{ritam.mallick5@gmail.com}
\affiliation{Frankfurt Institute for Advanced Studies, 60438 Frankfurt am Main, Germany}
%\and
\author{Stefan Schramm}
\email{schramm@th.physik.uni-frankfurt.de}
\affiliation{Frankfurt Institute for Advanced Studies, 60438 Frankfurt am Main, Germany}

\begin{abstract}
Magnetars are compact stars which are observationally determined to have very strong surface magnetic fields of the order of $10^{14}-10^{15}$G.
The centre of the star can potentially have a magnetic field several orders of magnitude larger. We study the effect of 
the field on the mass and shape of such a star. In general, we 
assume a non-uniform magnetic field inside the star which varies with density. The 
magnetic energy and pressure as well as the metric are expanded as multipoles in spherical harmonics up to the quadrupole term. 
Solving the Einstein equations for the gravitational potential, one obtains the correction terms as functions of the magnetic field. 
Using a nonlinear model for the hadronic EoS the excess mass and 
change in equatorial radius of the star due to the magnetic field are quite significant if the surface field is $10^{15}$G and 
the central field is about $10^{18}$ G.
For a value of the central magnetic field strength of $1.75\times10^{18}$ G, we find that both the excess mass and the equatorial radius 
of the star changes by about $3-4\%$ compared to the spherical solution.
\end{abstract}



\maketitle

\section{Introduction} \label{introduction}
Pulsars are among the most important ``laboratories'' to study the properties of 
matter at extreme conditions. They are known to emit waves of almost every 
wavelength, from x-rays to gamma rays. Connecting them with neutron stars (NS) \cite{gold} 
opened up a whole new branch of physics dealing with the equation of state (EoS)
of matter at extreme densities in connection with huge gravitational effects. The recent 
observational evidence of two solar mass neutron stars has generated significant additional activity in this field 
\cite{demorest,new2m}. The other
important feature of pulsars are the large surface magnetic fields. Usually, 
the observed surface magnetic field of pulsars ranges from $10^8-10^{12}$G. 
However, some new classes of pulsars, namely the anomalous X-ray pulsars (AXPs) 
and soft-gamma repeaters (SGR), have been identified to have much
higher surface magnetic field. The SGR are usually associated with supernova
remnants, which points to the fact that they are young NS \cite{kulkarni, 
murakami}. Recent measurements of the spin-down and the rate of change of spin-down suggest
that they are quite different from the bulk pulsar population with a surface 
magnetic field as high as $10^{15}$G. Observation of some X-ray pulsars also suggests that they can have 
surface fields of strengths of $10^{14}-10^{15}$G. The relation between the SGR and X-ray pulsars is
not quite clear, but we definitely have a class of NS with very high magnetic
fields, termed as magnetars \cite{duncan,thompson93,thompson95,thompson96}. 

The properties of NS, i.e., mass, radius, spin, etc., depend very sensitively 
on the EoS of matter describing the NS. However, in magnetars they also
depend sensitively on the magnetic field. Firstly, the matter in a strong 
magnetic background experiences two quantum effect, the Pauli 
paramagnetism (interaction of the spin of the fermion with the magnetic field) and 
Landau diamagnetism. Secondly, the magnetic pressure due to the Lorentz force 
induces a deformation of the star. The background magnetic 
field also affects the cooling and the magnetic field evolution 
of a neutron star. Hence, it is important to study the deformation of NS
in presence of strong magnetic fields.

The effect of a strong magnetic field on dense hadronic matter has been 
extensively studied in previous works \cite{chakrabarty,bandyopadhyay,broderick,
chen,rabhi}. The high magnetic field can affect the hydrostatic equilibrium
of NS and render the star unstable. The deformation of
magnetised NS was first discussed by Chandrasekhar \& Fermi and by Ferraro \cite{chandra,ferraro}.
The limiting field strength of the magnetic field was found to be of the order of 
$10^{18}$G. Instabilities related to the anisotropy of magnetic pressure was
also extensively discussed \cite{perez,huang,ferrer,dexheimer,sinha,sinha2}, both for
uniform and nonuniform magnetic fields. The anisotropy of the magnetic pressure in
the NS induces a deformation in NS, which we study in this paper. 

Calculations including the deformation of NS have been done before. 
The general relativistic approach by Bonazolla \& Gourgoulhon \cite{bonazzola} and 
by Bocquet et al. and by Cardall et al. \cite{bocquet,cardall} solves coupled matter and 
electromagnetic equations numerically. They solve exact GR equation for a dipolar magnetic field. Starting with a given 
current function and either a poloidal or a toroidal field they solve the
equations numerically. An analytic discussion was done by Konno et al. 
\cite{konno}, lacking a discussion involving a realistic EoS. 
In that work the field equations were treated perturbatively. 
Recent calculations \cite{ciolfi1,ciolfi2,ciolfi3}
also solve the stellar deformation due to magnetic fields following a perturbative approach.
In the latter articles the magnetic field is assumed to be of the form known as "twisted torus", where 
the field inside of the star, with a maximum value of $10^{16}$ G, is modelled as a combination of poloidal and toroidal fields, whereas
outside its form is simply poloidal.

In this article we follow a different reasoning. There is a large number of previous calculations, 
discussing the effect of the magnetic field on the EOS of the
neutron star. In most of these calculations \cite{chakrabarty,chakrabarty1,bandyopadhyay,mao,mallick,
sinha,dexheimer,lopes,dexheimer1} a density-dependent magnetic field profile with large central magnetic fields is assumed. Although the
magnetic pressure is anisotropic, in order to make the calculation more tractable, the magnetic pressure was isotropically added or 
subtracted to the total pressure.
This is not a correct approach, especially for large magnetic field values, and the problem needs a general 2D treatment. 
We take into account the anisotropic magnetic pressure and treat it as a perturbation similar to the method developed by
Hartle and Thorne \cite{hartle,thorne} for slowly rotating NS. We employ
a strong, density-dependent magnetic field distribution of frozen-in field with 
maximum strengths of the field in the core of the star of the order of $10^{18}$ G. 

The motivation of this work is to carry out the semi-analytic calculation of the 
deformation of a neutron star, caused by a non-uniform magnetic field 
pressure along different directions. We 
treat the non-uniform pressure as a perturbation to the total pressure (matter
and magnetic) and study its effect for the deformed star. In particular, we determine 
the excess mass and the ellipticity of the deformed star. We also comment on the possible
instability of a NS for a given field strength.

The paper is organised as follows. In Section II we carry out the calculation
of the deformation of the NS for anisotropic pressure up to the quadrupole
term. In section III we employ a realistic NS EoS and numerically 
calculate the excess mass and the ellipticity of the star, which yields the 
deformation of the star due to the magnetic effect. In Section IV we
summarise and discuss our results. 

\section{Formalism}\label{formalism}

In the rest frame of the fluid the magnetic field is aligned along the z-axis, and so the total energy density and
pressure takes the form 
\begin{eqnarray}
 \varepsilon=\varepsilon_m+\frac{B^2}{8\pi} \\
 P_{\perp}=P_m-MB+\frac{B^2}{8\pi} \\
 P_{\parallel}=P_m-\frac{B^2}{8\pi}.
\end{eqnarray}
where $\varepsilon$ is the total energy density, $\varepsilon_m$ is the matter energy density and $\frac{B^2}{8\pi}$
is the magnetic stress. $P_{\perp}$ and $P_{\parallel}$ are the perpendicular and parallel components
of the total pressure with respect to the magnetic field. $P_m$ is the matter pressure and $MB$ is the magnetization.
It has been discussed earlier in the literature that the effect of landau quantization on the EoS is negligible for reasonable magnetic 
fields \cite{sinha,dexheimer}.
The significant magnetic effect arises from the extra stress and pressure terms. Also, the effect due to magnetization 
is not very significant even for very strong fields, when the star itself becomes unstable due to very high magnetic fields at the centre. In our 
calculation we neglect 
all these less important effects and only deal with the magnetic stress and magnetic pressure. Therefore, the energy tensor can be written as 
\begin{eqnarray}
 \varepsilon=\varepsilon_m+\frac{B^2}{8\pi} \\
 P_{\perp}=P_m+\frac{B^2}{8\pi} \\
 P_{\parallel}=P_m-\frac{B^2}{8\pi}.
\end{eqnarray}
 
The pressure part is given as
\begin{eqnarray}
P=P_m\pm P_B \\
P=P_m+\frac{B^2}{8\pi}(1-2cos^2\theta).
\end{eqnarray}
where, $P_B$ is the magnetic pressure and $\theta$ is the polar angle with respect to the direction of the magnetic field. We can 
rewrite the total pressure as an expansion in spherical harmonics
\begin{eqnarray}
 P=P_m+\frac{B^2}{8\pi}[\frac{1}{3}-\frac{4}{3}P_2(cos\theta)] \\
 P=P_m+[p_0+p_2P_2(cos\theta)].
\end{eqnarray}
$p_0=\frac{B^2}{3.8\pi}$ is the monopole contribution and $p_2=-\frac{4B^2}{3.8\pi}$ the quadrupole contribution of the
magnetic pressure. $P_2(cos\theta)$ is the second order Legendre polynomial and is defined as 
\begin{eqnarray}
P_2(cos\theta)=\frac{1}{2}(3cos^2\theta -1).
\end{eqnarray}

We first assume that the neutron star is spherically symmetric. The interior solution of a static spherically 
symmetric object can be written in terms of Schwarzschild coordinates $t,r,\theta,\phi$ as
\begin{equation}
ds^2=-e^{\nu(r)}dt^2+e^{\lambda(r)}dr^2+r^2(d\theta^2+sin^2\theta d\phi^2),
\end{equation}
where the metric functions $\nu(r)$ and $\lambda(r)$ are function of $r$ only. The metric functions can be expressed as 
\begin{eqnarray}
\frac{d\nu}{dr}=-\frac{2}{\varepsilon_m+P_m}\frac{dP_m}{dr}, \\
e^{\lambda}=(1-\frac{2Gm(r)}{r})^{-1},
\end{eqnarray}
where $G$ is the gravitational constant and $m(r)$ is the mass enclosed in a sphere of radius $r$.

The general metric can also be formulated as a multipole expansion. In accordance with eqn. (9) and (10),
we only take along terms up to the quadrupole term. Hence, the metric can be written as \cite{hartle,chandra}
\begin{eqnarray}
ds^2=-e^{\nu(r)}[1+2(h_0(r)+h_2(r)P_2(cos\theta))]dt^2 \\
      +e^{\lambda(r)}[1+\frac{e^{\lambda(r)}}{r}(m_0(r)+m_2(r)P_2(cos\theta))]dr^2 \\
      +r^2[1+2k_2(r)P_2(cos\theta)](d\theta^2+sin^2\theta d\phi^2),
\end{eqnarray}
where $h_0,h_2,m_0,m_2,k_2$ are the corrections up to second order.

Solving the Einstein equations, we get
\begin{eqnarray}
\frac{dm_0}{dr}=4\pi r^2p_0 ,
\label{m0} \\
\frac{dh_0}{dr}=4\pi re^{\lambda}p_0+\frac{1}{r}\frac{d\nu}{dr}e^{\lambda}m_0+\frac{1}{r^2}e^{\lambda}m_0,
\label{h0} \\
\frac{dh_2}{dr}+\frac{dk_2}{dr}=h_2(\frac{1}{r}-\frac{\frac{d\nu}{dr}}{2})+\frac{e^{\lambda}}{r}m_2(\frac{1}{r}+\frac{{d\nu}{dr}}{2}),\\
h_2+\frac{e^{\lambda}}{r}m_2=0,
\label{m2} \\
\frac{dh_2}{dr}+\frac{dk_2}{dr}+\frac{1}{2}r\frac{d\nu}{dr}\frac{dk_2}{dr}=4\pi re^{\lambda}p_2+\frac{1}{r^2}e^{\lambda}m_2 \\ \nonumber
+\frac{1}{r}\frac{d\nu}{dr}e^{\lambda}m_2+\frac{3/r}e^{\lambda}h_2+\frac{2}{r}e^{\lambda}k_2.
\end{eqnarray}

From the conservation law of the total momentum, we obtain
\begin{eqnarray}
\frac{dp_0}{dr}=-\frac{d\nu}{dr}p_0-(\varepsilon+P)\frac{dh_0}{dr},\\
p_2=-(\varepsilon+P)h_2,\\
\frac{dp_2}{dr}=-\frac{d\nu}{dr}p_2-(\varepsilon+P)\frac{dh_2}{dr}.
\end{eqnarray}
With some simple algebra the fields can be expressed in terms of known quantities
\begin{eqnarray}
\frac{dk_2}{dr}=\frac{2p_2\frac{d\nu}{dr}+\frac{dp_2}{dr}}{\varepsilon+P},
\label{k2} \\
\frac{dh_2}{dr}=\frac{-p_2\frac{d\nu}{dr}-\frac{dp_2}{dr}}{\varepsilon+P}.
\label{h2}
\end{eqnarray}
Solving equations \ref{m0},\ref{h0},\ref{m2},\ref{k2},\ref{h2} for given values of $p_0$ and $p_2$, we can calculate $m_0,h_0,m_2,k_2$ and $h_2$.

The total mass of the star $M$ is given by
\begin{equation}
 M=M_0+\delta M,
\end{equation}
where $M_0$ is the mass for the matter part and $\delta M\equiv m_0$ is the additional mass due to the magnetic corrections. The shape of the star also 
gets deformed by the magnetic field, which is non-isotropic. The deformation can be quantified by the ellipticity 
($e$), which is defined as 
\begin{equation}
 e=\sqrt{1-\left(\frac{R_p}{R_e}\right)^2},
 \label{eccn}
\end{equation}
where $R_p$ is the polar radius and $R_e$ is the equatorial radius, respectively.
At this point all the metric perturbation potentials are specified. If we know the applied magnetic field and the initial matter perturbation
functions we can calculate the given metric perturbation potentials, and determine the mass change and deformation of the star. 

\section{Results}\label{results}

The static, spherically symmetric star can be solved using the TOV equation \cite{tov}. The pressure and enclosed mass of the star 
is given by
 \begin{eqnarray}
  \frac{dP(r)}{dr}&=&-\frac{G m(r) \varepsilon(r)}{r^2 } \,
  \frac{\left[ 1 + {P(r) / \varepsilon(r)} \right] 
  \left[ 1 + {4\pi r^3 P(r) / m(r)} \right]}
  {1 - {2G m(r)/ r} },
\\
  \frac{dm(r)}{dr}&=&4 \pi r^2 \varepsilon(r). \\
 \end{eqnarray} 
 The total mass is defined as
 \begin{eqnarray}
  M_G~ \equiv ~ m(R)  = 4\pi \int_0^Rdr~ r^2 \varepsilon(r),
\end{eqnarray}
where $R$ is the radius of the star. 
The TOV equation is solved for a given central energy density corresponding to a central pressure. The surface of the neutron star
$r=R$, is defined as the point where the pressure vanishes. Along with this solution, 
we also solve for the expanded perturbation potentials for a given EoS and given magnetic field profile. In our problem we show results 
for two different hadronic EoS. The assumed magnetic profile of the star is assumed to be density dependent 
\cite{chakrabarty,chakrabarty1,bandyopadhyay,mao,mallick,sinha,dexheimer,lopes,dexheimer1}, and is parametrized as
\begin{equation}
{B}(n_b)={B}_s+B_0\left\{1-e^{-\alpha \left(\frac {n_b}{n_0} \right)^\gamma}\right\}.
\label{mag-vary}
\end{equation}
This simple ansatz covers a general physical situation where the magnetic field in the star is non-uniform. The model is constructed
in such a way that the magnetic field at the centre of the star can in principle be several orders of magnitude larger than at the surface. 
The parameters $\alpha$ and $\gamma$ 
control how fast the central magnetic field $B_c$ falls to the asymptotic value at the surface $B_s$. 
Observationally the surface magnetic field strength of magnetars are usually of the order
of $10^{14}-10^{15}$G. The central magnetic field strength might be as high as $10^{18} -10^{19}$ G, e.g. assuming some dynamo effect inside the star 
as discussed in 
\cite{duncan}. For an asymptotic field of $B_0=4\times10^{18}$ G, the value of the magnetic field at the centre of the star (whose central density
is not greater than $8$ times nuclear saturation density) is not greater than $B_c=1.75\times10^{18}$ G. Effectively, this is the maximum value of
the central magnetic field we are considering. With this value the ratio of magnetic pressure ($P_B$) to that of matter pressure
($P_m$) is always less than $0.5$, and the ratio of total magnetic energy to gravitational energy is less than $0.1$.
With such a choice of magnetic field strength, the perturbative calculation is expected to provide us with reasonable results.
In the case of NS rotation, the perturbative approach has shown to provide quantitatively good results by \cite{weber,glen}
even for millisecond pulsars. They differ slightly from exact GR calculations only if the star rotational velocity approaches keplerian
velocity. Therefore, in our problem, with $B_c$ in the range of $10^{17}-10^{18}$ G, our approach should still be within the range 
of its validity and certainly far better than the treatment used in Refs. \cite{mallick,dexheimer,lopes,dexheimer1}.
We keep the surface value of the magnetic field fixed at $10^{15}$G. We assume $\alpha=0.01$ and $\gamma=2$, which is quite a gentle 
variation of the magnetic field 
inside the star. Other $\alpha,\gamma$ combination yield different possible variations. However, the results for our calculation would not be
much affected and the qualitative conclusions would remain the same. 

\begin{figure}[ht]
\vskip 0.2in
\subfloat[]{\includegraphics[width = 3.1in]{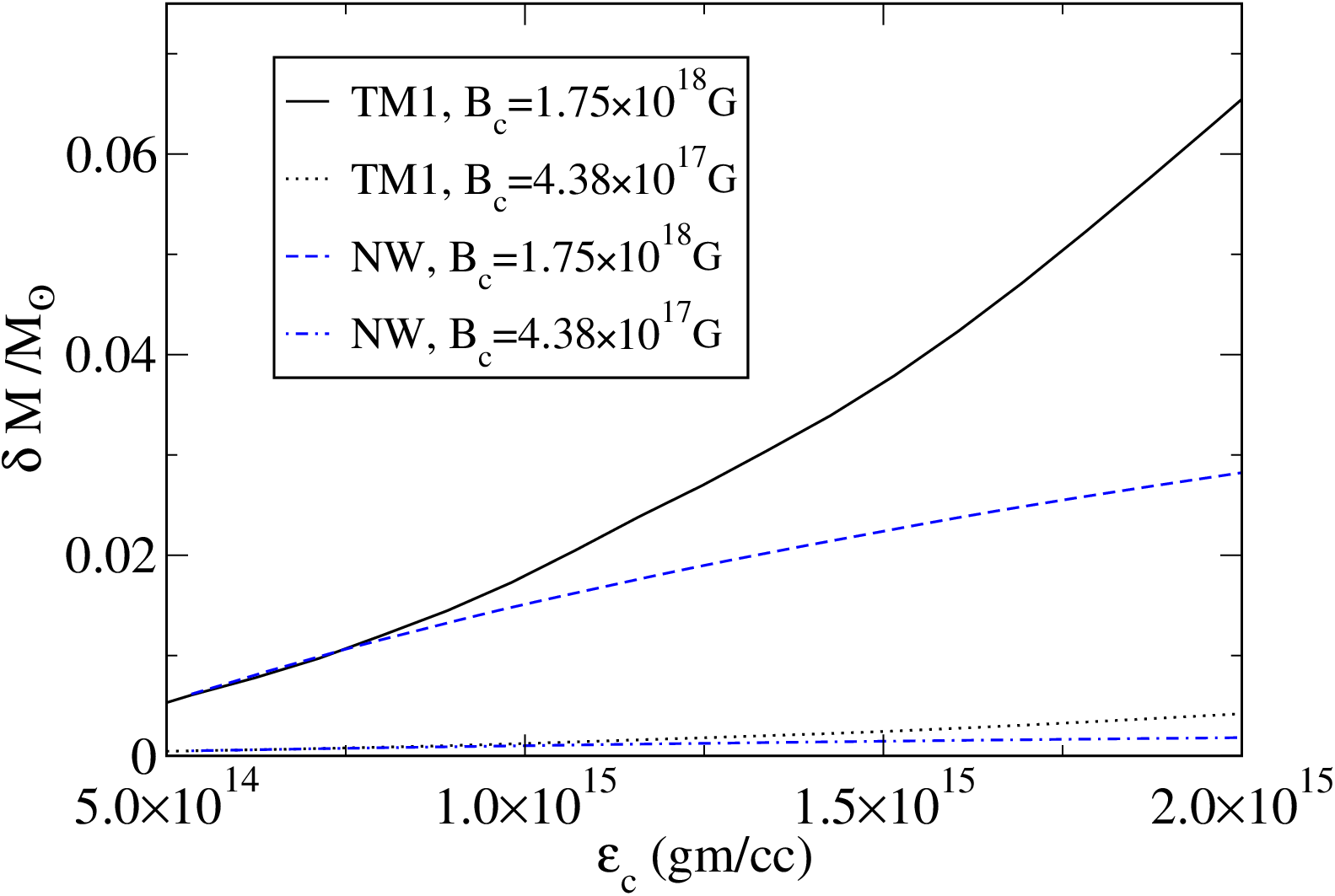}} \quad 
\subfloat[]{\includegraphics[width = 3.1in]{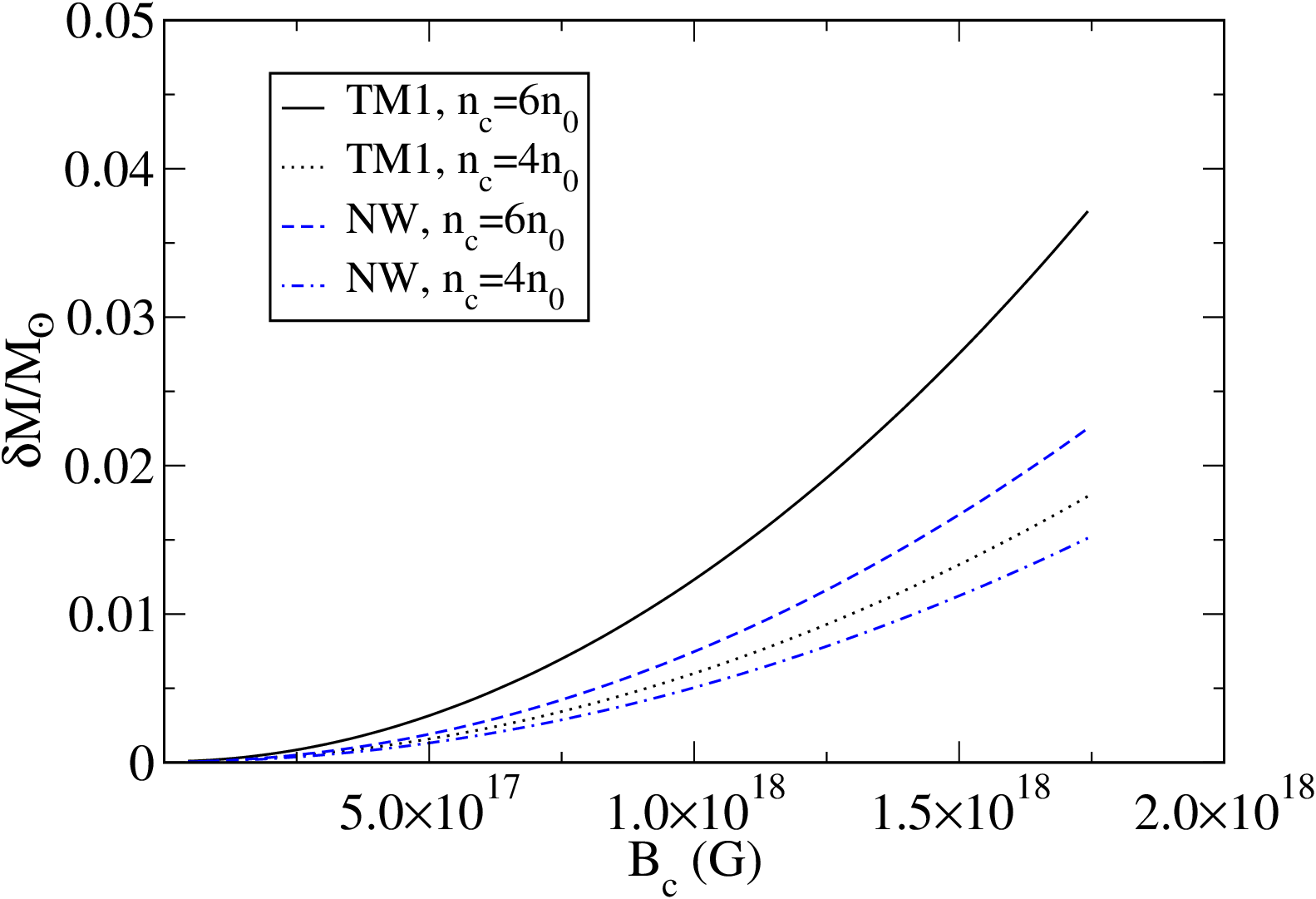}}
\caption{(Color online) $\delta M$ as a function of central energy density $\varepsilon_c$ for 
fixed $B_c$ (1a) and as a function of central magnetic field $B_c$ for fixed central density $n_c$ (1b). 
Curves are plotted for two different EoS (TM1 and NW). The central 
magnetic field $B_c$ and central density $n_c$ are specified in the figures.}
\label{fig1}
\end{figure}

\begin{figure}[ht]
\vskip 0.2in
\subfloat[]{\includegraphics[width = 3.1in]{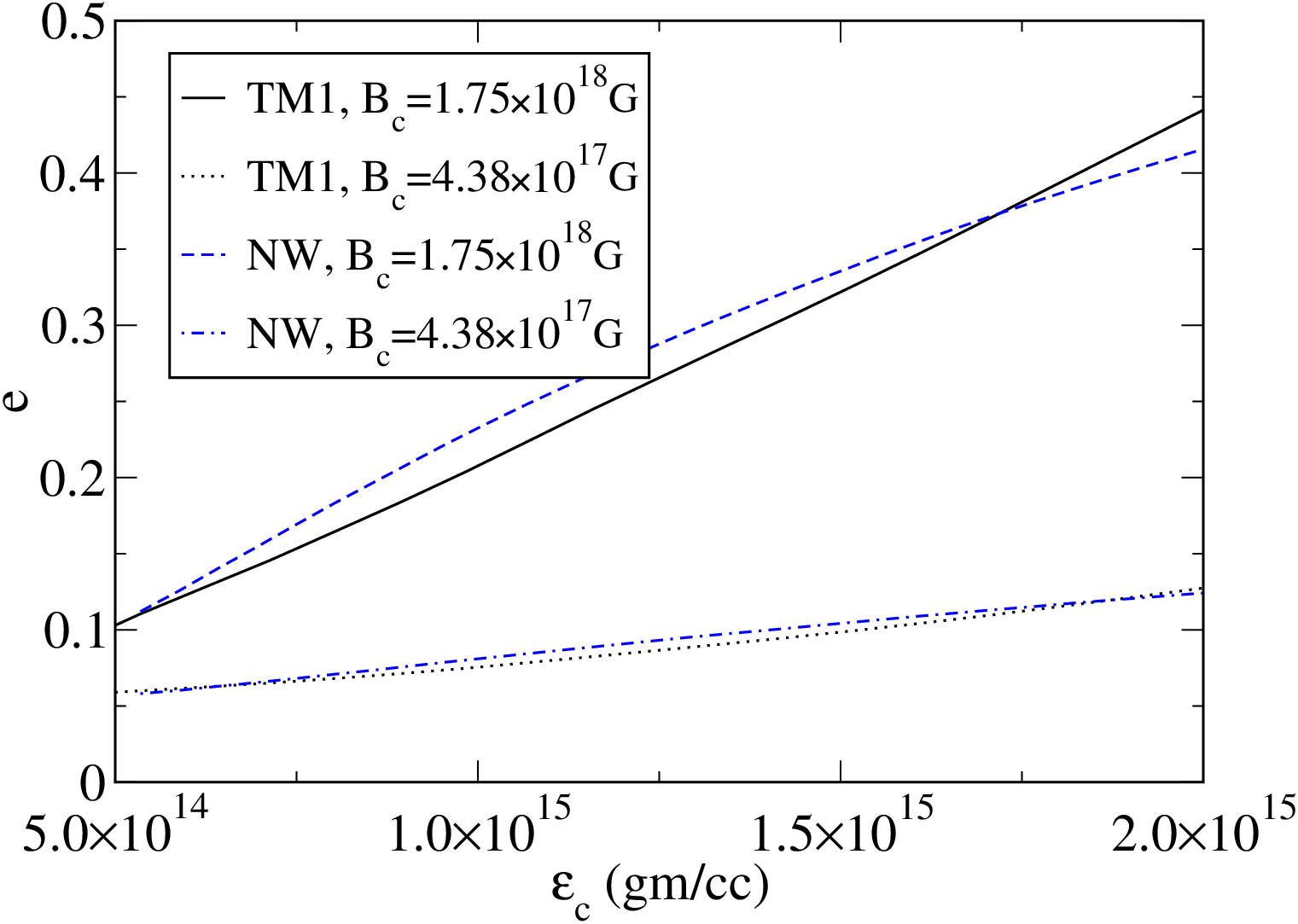}} \quad 
\subfloat[]{\includegraphics[width = 3.1in]{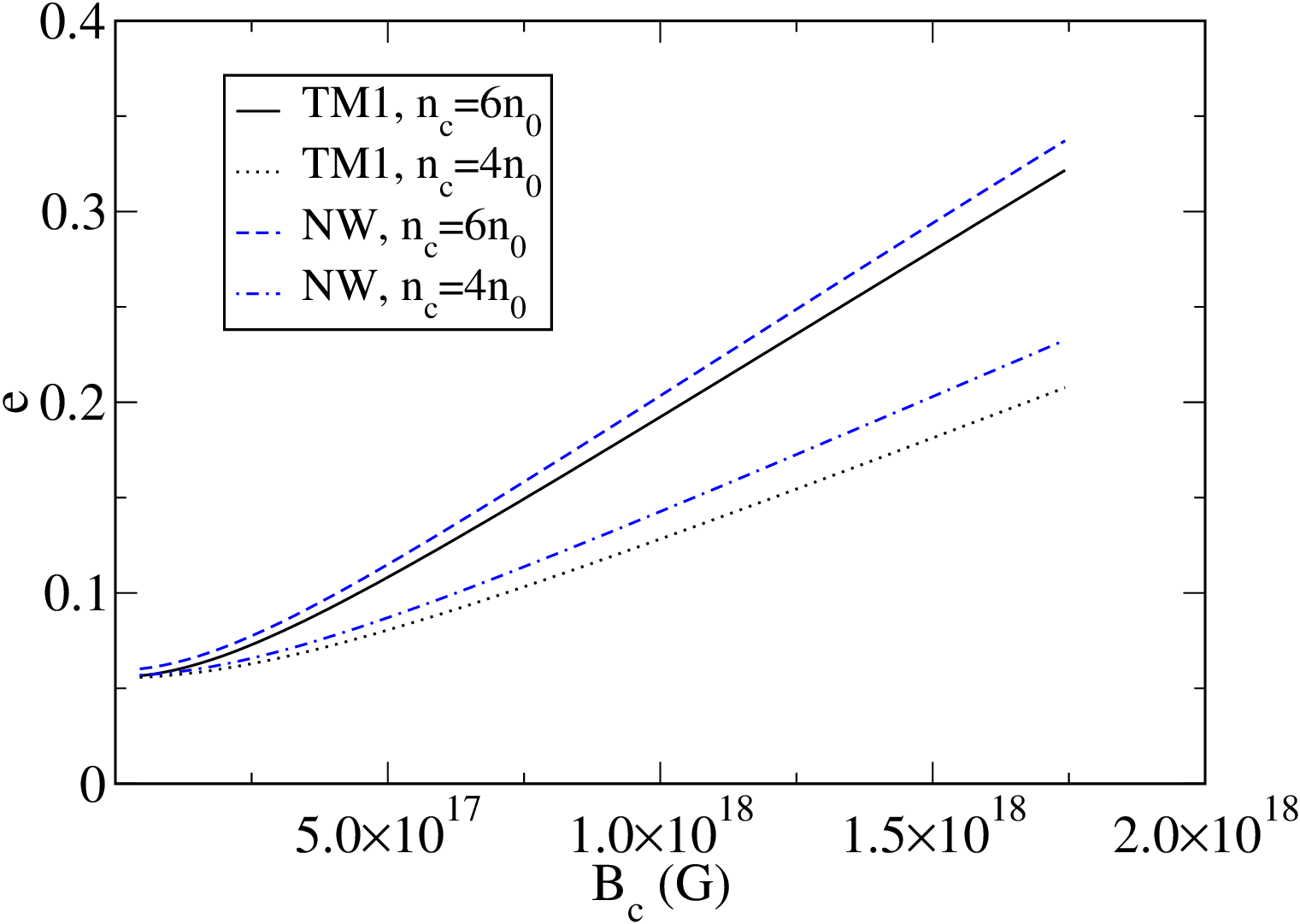}}
\caption{(Color online) The eccentricity $e$ as a function of $\varepsilon_c$ for fixed $B_c$ and as a function of $B_c$ for fixed $n_c$ is plotted. 
For comparison we have plotted curves for two different EoS.}
\label{fig2}
\end{figure}

For the two EoS, we choose a very stiff nuclear EoS, namely the nonlinear Walecka model \cite{walecka}, which is able to reproduce the mass of 
the observed pulsar PSR J1614-2230 \cite{demorest}. For comparison we also use a much softer EoS (TM1) \cite{sugahara,schaffner} that includes
hyperons. 

Fig. 1a and 1b show the excess mass of the star due to the magnetic field. The excess mass is related to
the $m_0$ component of the correction for the monopole term ($\delta M=m_0(R)$). 
Figure 1a shows the variation of extra mass with central energy density
($\varepsilon_c$). We have plotted curves for two different 
central magnetic fields. As the central energy density increases, the excess mass also increases, 
irrespective of the value of $B_c$.
This is a direct result of Eqn.~\ref{mag-vary}, as with increasing central energy density the corresponding 
number density and therefore the central magnetic field become larger as well.
 For a field of $B_c=1.75\times10^{18}$ G the the excess mass if of the order of few percent (maximum for 
TM1 parametrization with $3-4\%$). As the magnetic field decreases to $B_c=4.38\times10^{17}$ G the excess mass 
becomes one order of magnitude less.
Figure 1b shows the variation of excess mass with central magnetic field $B_c$. The curves are plotted for fixed central density $n_c$ 
of $4$ and $6$ times nuclear saturation density ($n_0$), also corresponding to a fixed central energy density.
As expected the excess mass increases with an increase in magnetic field.
For the stiff EoS the excess mass is less than for the softer EoS, because the ratio of the magnetic pressure to
matter pressure is smaller for a stiffer EoS. 

As the anisotropic magnetic pressure generates excess mass for the star, it is likely that it also produces a significant 
deformation. The magnetic pressure adds to the matter pressure in the equatorial 
direction and reduces it along the polar direction. Therefore, we expect a flattening of the star, taking a shape of an 
oblate spheroid (similar to the deformation due to rotation). 
The polar and equatorial radii of a deformed star are defined as 
\begin{eqnarray}
 R_e=R+\xi_0(R)-\frac{1}{2}(\xi_2(R)+rk_2), \\
 R_p=R+\xi_0(R)+(\xi_2(R)+rk_2),
\end{eqnarray}
where $R$ is the radius of the spherical star. $\xi_0$ and $\xi_2$ are defined as
\begin{eqnarray}
 \xi_0(r)=\frac{r(r-2Gm(r))}{G(4\pi r^3P_m+m(r))}p_0^*, \\
 \xi_2(r)=\frac{r(r-2Gm(r))}{G(4\pi r^3P_m+m(r))}p_2^*,
\end{eqnarray}
with $p_0^*$ and $p_2^*$ given by
\begin{eqnarray}
p_0^*=p_0/\frac{B^2}{8\pi}, \\
p_2^*=p_2/\frac{B^2}{8\pi}.
\end{eqnarray}

Thus, the polar and equatorial radii of a star have contributions from the three terms $\xi_0, \xi_2$ and $k_2$. The contribution of the 
$\xi$'s originates from the surface magnetic field strength of the magnetar, and $k_2$ is the contribution from the integrated magnetic pressure 
throughout the star. The deformation of the star is given by the deformation parameter called eccentricity $e$ as defined in eqn. \ref{eccn}.
Fig. 2a studies the variation of $e$ with central energy density. Eccentricity increases with increase in central energy density and is 
more or less same for both EOS. For central magnetic field of
strength $B_c=1.75\times10^{18}$ G, $e$ varies in the range $0.1-0.5$, and with lower central field strength
($B_c=4.38\times10^{17}$ G) it is about half of the previous value. As the central magnetic field increases, the magnetic 
pressure contribution also rises and thereby the deformation of the star. We also show $e$ as a function of central magnetic 
field ($B_c$) (Fig. 2b) for fixed central energy density ($n_c$). The eccentricity is an increasing function of central magnetic field as shown 
in the figure.

\begin{figure}[ht]
\vskip 0.2in
\subfloat[]{\includegraphics[width = 3.1in]{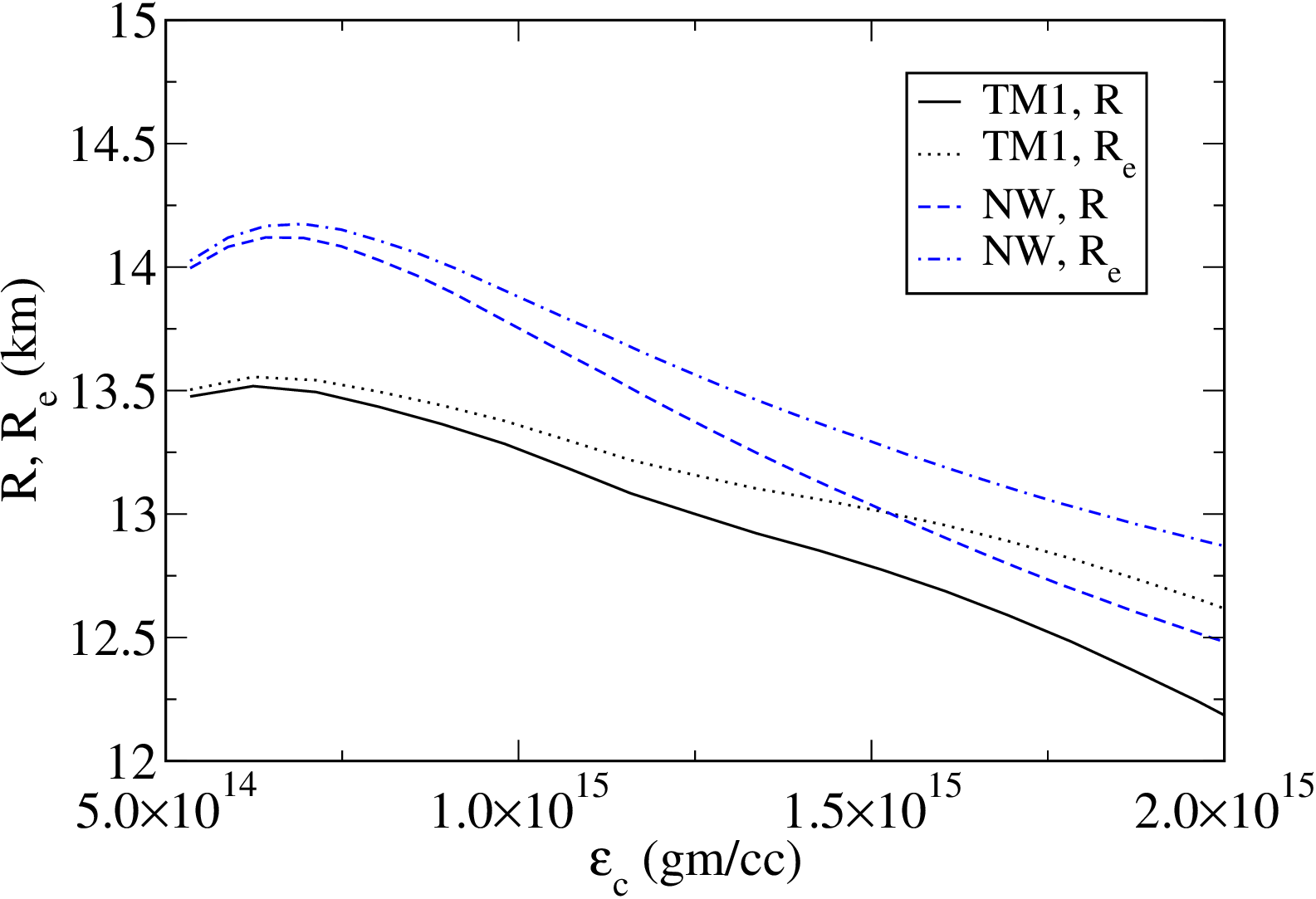}} \quad 
\subfloat[]{\includegraphics[width = 3.1in]{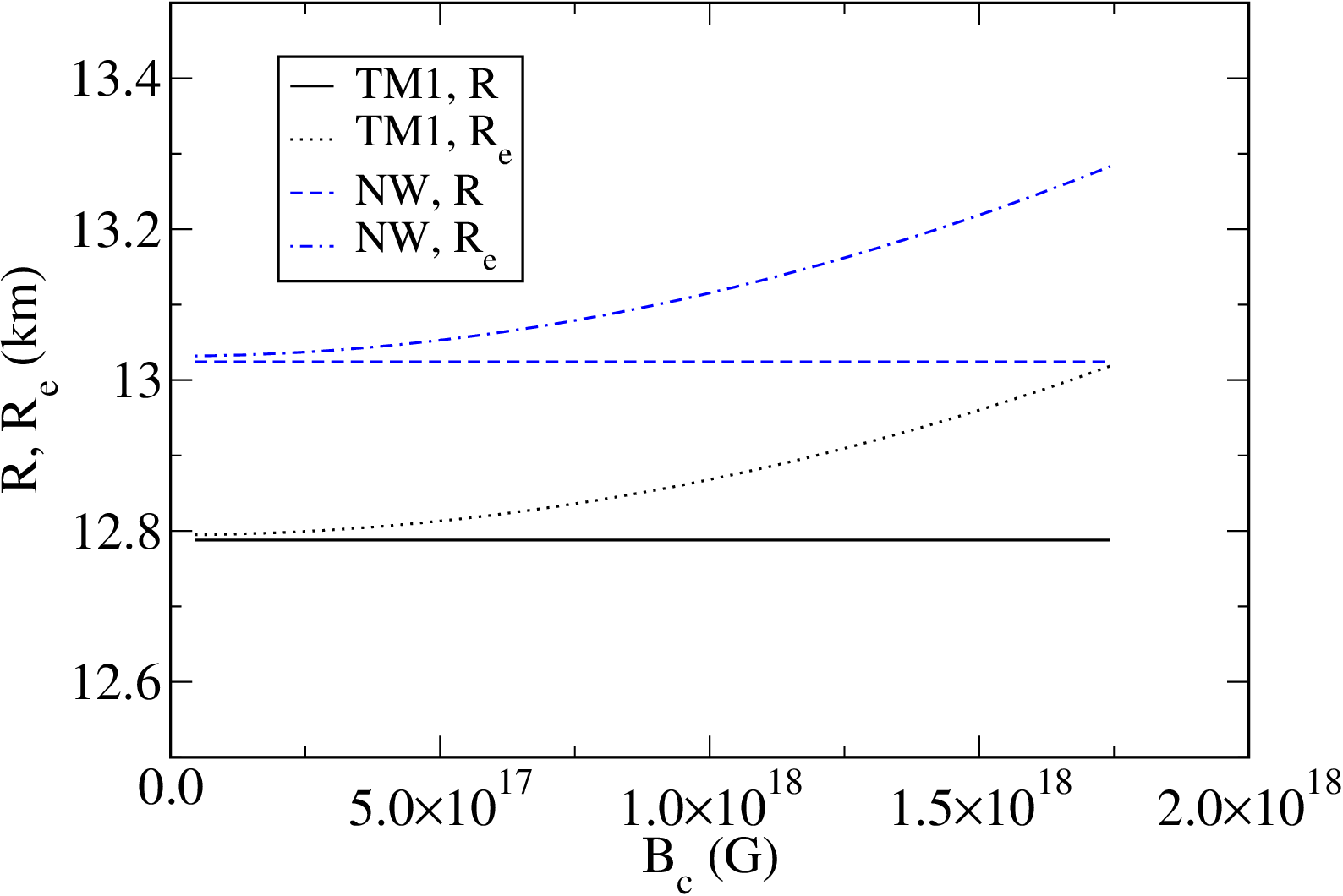}}
\caption{(Color online) Radii $R_e,R$ plotted as functions of central energy density (for $B_c=1.75\times10^{18}$ G) and as a 
function of central magnetic field (for $n_c=6n_0$). Curves for two different EoS, TM1 and NW model are shown.}
\label{fig3}
\end{figure}

\begin{figure}[ht]
\vskip 0.2in
\subfloat[]{\includegraphics[width = 3.1in]{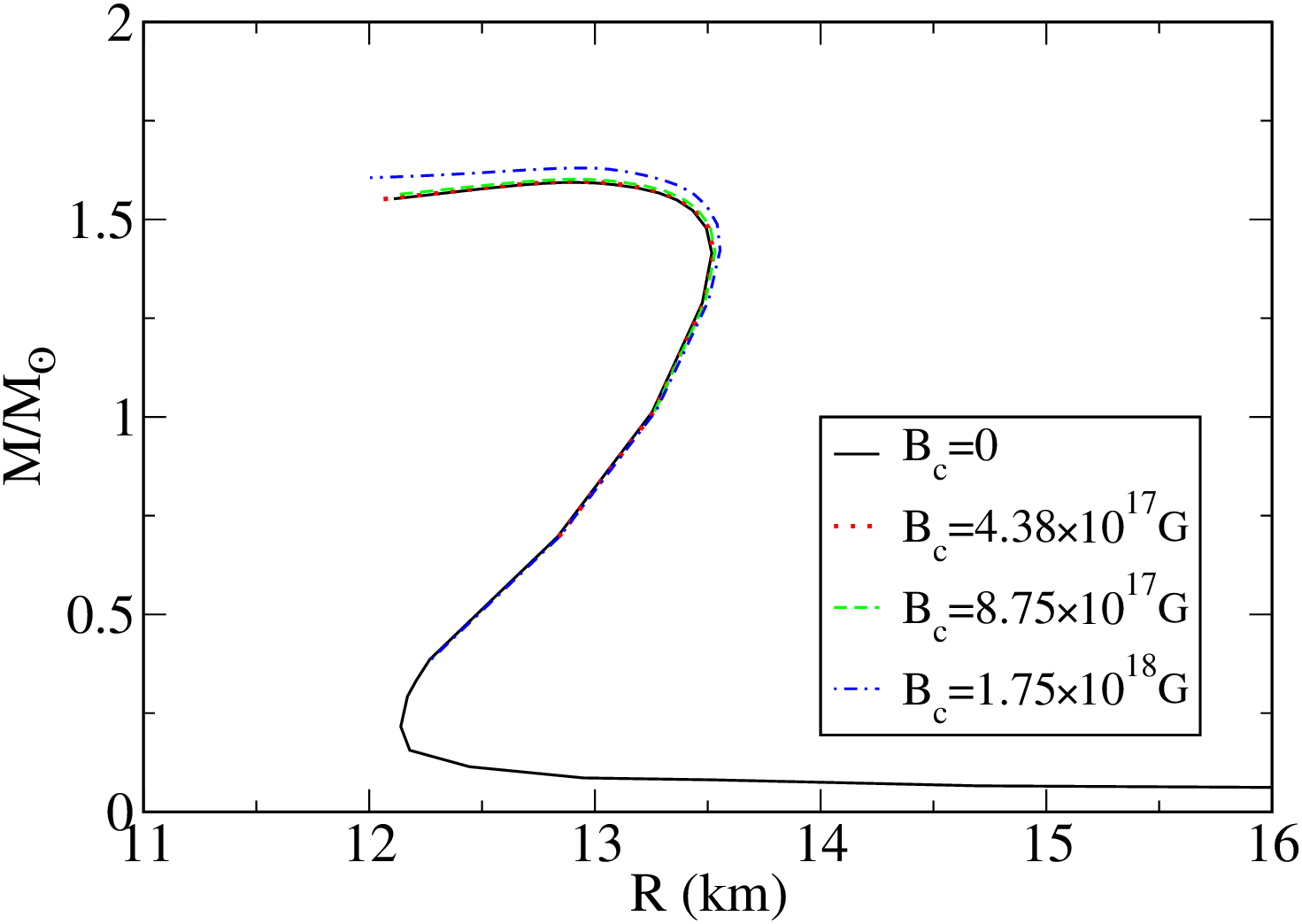}} \quad 
\subfloat[]{\includegraphics[width = 3.1in]{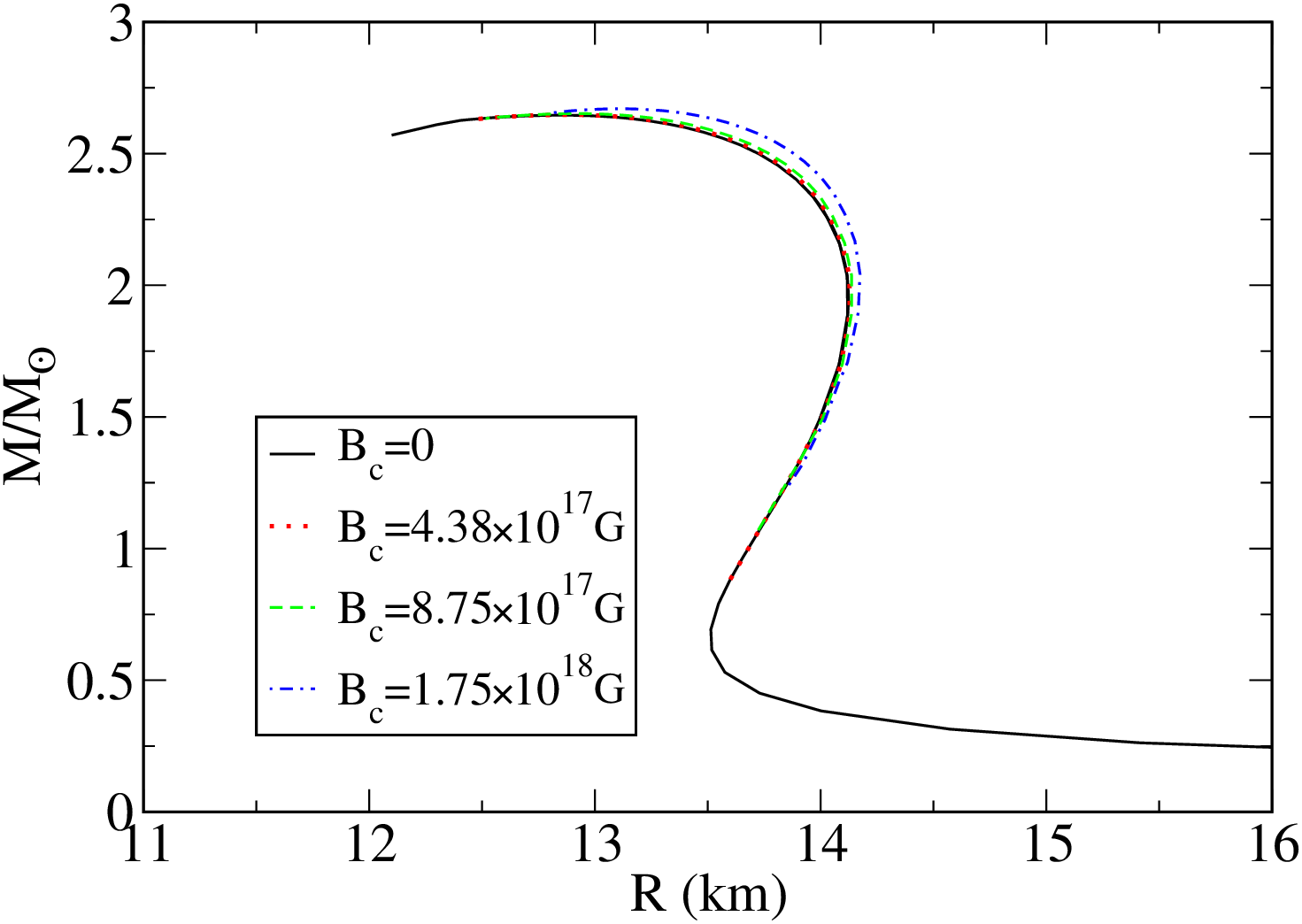}}
\caption{(Color online) Mass-radius curve for TM1 and NW models.}
\label{fig4}
\end{figure}

To see how the star radius changes with magnetic field, we show the difference of the radius of the spherical star (without magnetic field)
and the equatorial radius of the deformed star with energy density and magnetic field. Fig. 3a presents the radial variation with 
central energy density. Curves are for magnetic field of $B_c=1.75\times10^{18}$ G are shown. As the central energy density increases the 
radius $R$ becomes smaller. However, $R_e$, which also has positive contribution from the magnetic field does not decrease as much as $R$, and for any 
given central density $R_e > R$. Along the polar direction the picture is completely different as the magnetic pressure has a negative contribution
and for any given central density $R_p < R$. This means that for a magnetized NS  we naturally obtain $R_e > R_p$ 
and the amount of inequality depends on the given central 
magnetic field strength. Fig. 3b illustrates this point more clearly. For a central energy density of $6n_0$, the radius of the star $R$ is constant
($\sim 13 km$), but $R_e$ increases together with the central magnetic field strength, generating a oblate shape. 

Finally, we consider the mass-radius curve of the star (Fig. 4a and 4b). We find that as the 
magnetic field increases the mass-radius curve becomes stiffer and the maximum mass of the star increases. As one might expect, for the softer 
EOS the maximum is  increased by a larger extent compared to the stiffer EOS. The mass-radius curve also provides a hint of the maximum central 
magnetic field that can be present inside the star. As the magnetic field increases, the mass-radius curve at maximum value becomes increasingly 
flat (plateau like) with respect to a change of the radius. If we increase the magnetic field further ($B_c=2\times10^{18}$ G), there is no 
maximum mass any more , and the mass radius 
curve goes on increasing at the lower radii region of the plot. This is due to the fact that excess mass from the magnetic field dominates
the mass drop due to the matter counterpart. Although the overall magnetic energy is less than the gravitational energy
of the star, the magnetic pressure at the centre approaches the matter pressure and generates this peculiar feature, which clearly 
signals a limit of the applicability of the outlined approach.

Static stars, as they are discussed in this article, do not emit gravitational waves.
However a rotating star, which deviates from axisymmetry, does. Therefore, an estimate of the strength of gravitational wave (GW) emission 
can be deduced if we assume that the magnetic axis and the rotation axis are 
not aligned, as it is the case in observable pulsars. Let us make some crude estimate of such a GW strength from some well-known magnetar candidate.
The amplitude of the GW signal is given by \cite{bonazzola2,ciolfi2}
\begin{equation}
 h_0= \frac{4 G}{d c^4} \Omega^2 I \epsilon \sin\alpha
\end{equation}
where, $d$ is the distance of the magnetar, $c$ the speed of light, and $\Omega$ the rotational velocity of the magnetar, $I$ is the moment of 
inertia with respect to the rotational axis, $\epsilon$ denotes the ellipticity of the star and $\alpha$ is the angle between the rotation and 
magnetic axis. The ellipticity 
of the star is related to its eccentricity by \cite{ferraro}
\begin{equation}
 \epsilon=\frac{1}{2}e^2.
\end{equation}
If we assume the mass of the star to be around $1.5$ solar masses, we can determine its moment of inertia and ellipticity from our calculation. 
Taking the case of the magnetar SGR 1900+14 (whose surface magnetic field strength is about $7\times10^14$ G), the period of rotation is $5$ sec 
and $d$ is $12.5$ kpc. The moment of inertia of a star of mass $1.5$ solar mass (having maximum possible magnetic field) with the stiff $NW$ EoS is 
$2.35\times10^{38}$ $kg m^2$ and the ellipticity is $0.006$.
With these values the amplitude of the gravitational wave $h_0$ is calculated to be $2.41\times10^{-28} \sin\alpha$. As the absolute value of 
$\sin\alpha$ cannot exceed $1$, 
we always obtain a value for $h_0$ of the order of $10^{-28}$. With the TM1 parameter set the result is similar ($\epsilon=0.0098$ and 
$h_0=3.58\times10^{-28}$).
Such estimate with SGR 10501+4516 (which is much closer to us, $d=2$ kpc) leads to $h_0=4.65 \times 10^{-28} \sin\alpha$ and with 
AXP 1E 1841-045 to $h_0=4.54 \times 10^{-29} \sin\alpha$.
Therefore, the expected maximum GW strength is likely to be about $10^{-28}$ for nearly all magnetars.
Given the capabilities of the VIRGO detector stated for a frequency of $30$ Hz \cite{bonazzola3},
the minimum amplitude detectable within three years of data integration is $h_{min} \sim 10^{-26}$. Therefore, the result is not very encouraging.
However, the situation may drastically change if we have a magnetar much closer to us (around $2$ kpc), and particularly with a significantly 
reduced rotational period (for example, about $100$ ms). 
Such magnetar would lead to GW with amplitudes of the order of $h_0 \sim 10^{-23}-10^{-24}$, which would clearly be in the range of detectability 
with the VIRGO and LIGO detectors.

\section{Summary \& Discussion}\label{summary}

In the present work we have carried out a semi-analytic calculation of the deformation of a neutron star assuming non-uniform magnetic pressure along 
different directions (equatorial and polar). We have treated the magnetic pressure as a perturbation to the total pressure. 
In general, we have assumed a non-uniform magnetic field distribution inside the star as was discussed in a number of papers 
\cite{chakrabarty,sinha,dexheimer,dexheimer1}.  We have
neglected the effect due to the magnetization of matter and the modification of the nuclear EOS due to the fields, because its contributions even at 
large magnetic fields is very small \cite{dexheimer}. We have expanded both 
the  pressure and energy density in spherical harmonics up to the quadrupole term. Analogously, 
we have also expanded the space-time metric, following similar approaches by previous authors \cite{hartle,chandra,konno}. 
Subsequently, we have solved the 
Einstein equations and obtained all the metric corrections as functions of known magnetic pressure contributions. 
Much more numerically involved
calculations have been done before \cite{bonazzola,bocquet,cardall}, however, our semi-analytical approach provides an 
intuitive and practical description  
of the excess mass and deformation of a star due to magnetic field effects, significantly improving the treatment 
for the magnetic field used previously.
 
We have solved the metric corrections for a given central and surface magnetic field. The correction terms are related both to the excess mass 
and deformation of the star. The monopole correction term $m_0$ yields the excess mass and the quadrupole correction term $k_2$ along with the 
surface magnetic field determines the deformation of the star. As expected the correction terms and the excess mass and deformation are 
proportional to the central and surface magnetic fields. The variation of the magnetic field inside the star affects both the mass and 
deformation, but only by a small amount.
The excess mass of the star due to the magnetic field adds to about $3-4\%$ of the original mass and the change in the equatorial 
radius of the star is also
about the same amount (which is quite different from previous $1D$ calculations which predicts ($10-15\%$ mass change) \cite{lopes,dexheimer1}). 
We have obtained a central bound on the magnetic field within this approach from the mass-radius diagram, 
beyond which the star fails 
to produce a maximum mass. 
As a practical limit of this approach we have assumed that the central field is always such that the magnetic 
energy to gravitational energy of the star is $< 0.1$. With the given EOS, this yields a central magnetic field close to 
$1.75\times10^{18}$ G. 

Fo a rotating star with nonaligned magnetic and rotational axes, the estimated 
GW strength for known magnetars is around $10^{-28}$, which is much less than the minimal amplitude expected to be detectable in LIGO and VIRGO 
interferometric detectors.
The situation can significantly improve by several orders of magnitude, in case magnetars with higher rotational frequency, than measured so far, exist. 

Note that so far we have not assumed any electric field or current distribution in our calculation. 
Also, the inclusion of the rotational effect would generate a 
finite electric field, further complicating the equations. However, this is an interesting scenario, as then
the rotational deformation adds to the magnetic one, limiting further the central magnetic field. Extended calculations along this line 
are in progress.

\begin{acknowledgments}
The authors would like to thank HIC for FAIR for providing financial support to the project.
\end{acknowledgments}

{}

%\clearpage


\begin{thebibliography}{}
\bibitem{gold} Gold, T., Nature, 218, 731 (1968)
\bibitem{demorest} Demorest, P., Pennucci, T., Ransom, S., Roberts, M., \& Hessels, J., Nature, 467, 1081 (2010)
\bibitem{new2m} Antonidis, J., Freire, P. C. C., Wex, N. et. al., Science, 340, 448 (2013) 
\bibitem{kulkarni} Kulkarni S. R., \& Frail, D. A., Nature, 365, 33 (1993)
\bibitem{murakami} Murakami, T., Tanaka, Y., Kulkarni, S. R., Ogasaka, Y., Sonobe, T., Ogawara, Y., 
Aoki, T., \& Yoshida, A., Nature,  368, 127 (1994)
\bibitem{duncan} Duncan, R. C., \& Thompson, C., AstroPhys. J., 392, L9 (1992)
\bibitem{thompson93} Thompson, C., \& Duncan, R. C., AstroPhys. J., 408, 194 (1993)
\bibitem{thompson95} Thompson, C., \& Duncan, R. C., Mon. Not. Roy. Astron. Soc., 275, 255 (1995)
\bibitem{thompson96} Thompson, C., \& Duncan, R. C., AstroPhys. J., 473, 322 (1996)
\bibitem{chakrabarty} Chakrabarty, S., Bandyopadhyay, D., \& Pal, S., Phys. Rev. Lett., 78, 2898 (1997)
\bibitem{bandyopadhyay} Bandyopadhyay, D., Chakrabarty, S., Dey, P., \& Pal, S., Phys. Rev. D, 58, 121301 (1998)
\bibitem{broderick} Broderick, A., Prakash, M., \& Lattimer, J. M., AstroPhys. J., 537, 351 (2000)
\bibitem{chen} Chen, W., Zhang, P. Q., \& Liu, L. G., Mod. Phys. Lett. A, 22, 623 (2005)
\bibitem{rabhi} Rabhi, A., Providencia, C., \& Providencia, J. Da, J. Phys. G Nucl. Phys., 35, 125201 (2008)
\bibitem{chandra} Chandrasekhar, S., \& Fermi, E., AstroPhys. J., 118, 116 (1953) 
\bibitem{ferraro} Ferraro, V. C. A., AstroPhys. J., 119, 407 (1954)
\bibitem{perez} Perez Martinez, A., Perez Rojas, H., \& Mosquera Cueata, H. J., Internat. J. Mod. Phys. D, 17, 2107 (2008)
\bibitem{huang} Huang, X. G., Huang, M., Rischke, D. H., \& Sedrakian, A., Phys. Rev. D, 81, 045015 (2010)
\bibitem{ferrer} Ferrer, E. J., delaIncera, V., Keith, J. P., Portillo, I., Springsteen, P. L., Phys. Rev. C, 82, 065802 (2010)
\bibitem{dexheimer} Dexheimer, V., Negreiros, R., \& Schramm, S., Eur. Phys. J. A, 48, 189 (2012)
\bibitem{sinha} Sinha, M., Mukhopadhyay, B., \& Sedrakian, A., Nucl. Phys. A, 898, 43 (2013)
\bibitem{sinha2} Sinha, M., Huang, X. G., Sedrakian, A., Phys. Rev. D, 88, 025008 (2013)
\bibitem{bonazzola} Bonazzola, S., Gourgoulhon, E., Salgado, M., \& Marck, J. A., Astron. \& AstroPhys., 278, 421 (1993)
\bibitem{bocquet} Bocquet, M., Bonazzola, S., Gourgoulhon, E., \& Novak, J., Astron. \& AstroPhys., 301, 757 (1995)
\bibitem{cardall} Cardall, C. Y., Prakash, M., \& Lattimer, J. M., AstroPhys. J., 554, 322 (2001)
\bibitem{konno} Konno, K., Obata, T., \& Kojima, Y., Astron. \& AstroPhys., 352, 211 (1999)
\bibitem{ciolfi1} Ciolfi, R., Ferrari, V., Gualtieri, L., \& Pons, J. A., Mon. Not. Roy. Astron. Soc., 397, 913 (2009)
\bibitem{ciolfi2} Ciolfi, R., Ferrari, V., \& Gualtieri, L., Mon. Not. Roy. Astron. Soc., 406, 2540 (2010)
\bibitem{ciolfi3} Ciolfi, R., \& Rezzolla, L., Mon. Not. Roy. Astron. Soc. lett., 435, L43 (2013)
\bibitem{chakrabarty1} Bandyopadhyay, D., Chakrabarty, S., \& Pal, S., Phys. Rev. Lett., 79, 2176 (1997) 
\bibitem{mao} Mao, G. J., Iwamoto, A., \& Zhu-Yia, Li, Chinese J. of Astron. \& AstroPhys., 3, 359 (2003)
\bibitem{mallick} Mallick, R., \& Sinha, M., Mon. Not. Roy. Astron. Soc., 414, 2702 (2011)
\bibitem{lopes} Lopes, L. L. \& Menezes, D. P., Braz. J. Phys., 42, 428 (2012)
\bibitem{dexheimer1} Dexheimer, V., Menezes, D. P., \& Strickland, M., J. Phys. G 41, 015203 (2014)
\bibitem{hartle} Hartle, J. B., AstroPhys. J., 150, 1005 (1967) 
\bibitem{thorne} Hartle, J. B., \& Thorne, K. S., AstroPhys. J., 153, 807 (1968)
\bibitem{tov} Shapiro S. L., \&  Teukolsky S. A., {\it Black Holes, White Dwarfs, and
Neutron Stars}, (John Wiley \& Sons, New York, 1983)
\bibitem{weber} Weber, F., {\it Pulsar as an astrophysical laboratory for nuclear and particle physics}, (Institute of 
Physics Publishing, Bristol, 1999)
\bibitem{glen} Glendenning, N. K., {\it Compact Stars: Nuclear Physics, Particle Physics, and General 
Relativity}, (Springer, New York, 2000)
\bibitem{walecka} Walecka, J. D., AstroPhys. J., 83, 491 (1974)
\bibitem{sugahara} Sugahara, Y., \& Toki, H., Nucl. Phys. A, 579, 557 (1994)
\bibitem{schaffner} Schaffner, J., \& Mishustin, I. N., Phys. Rev. C, 53, 1416 (1996)
\bibitem{bonazzola2} Bonazzola, S., \& Gourgoulhon, E., Astron. \& AstroPhys., 312, 675 (1996)
\bibitem{bonazzola3} Bonazzola, S., \& Marck, J. A., Annu. Re. Nucl. Part. Sci., 45, 655 (1994)
\end{thebibliography}
\end{document}